\documentclass{PoS}

\title{The anomalous dimension at the infrared fixed point of $N_f = 12$ SU($3$) theory}

\ShortTitle{The anomalous dimension at the IRFP of $N_f = 12$ SU(3) theory}

\author{\speaker{Etsuko Itou}\\
        High Energy Accelerator Research Organization (KEK), Tsukuba 305-0801, Japan\\
        E-mail: \email{eitou@post.kek.jp}}

\abstract{We propose a novel renormalization scheme for the hadronic operators.
As an example, we show the numerical simulation result for the anomalous dimension of the pseudo scalar operator of the SU($3$) gauge theory coupled to $N_f = 12$ fundamental fermions.
The anomalous dimension of the pseudo scalar operator is related with the mass renormalization factor of the fermion thought the partially conserved axial-vector current
(PCAC) relation.
We derive the mass anomalous dimension at the infrared fixed point (IRFP) of the theory, and find that it is given by
$\gamma_m^*= 0.044 \hspace{5pt}_{-0.024}^{+0.025} (\mbox{stat.}) _{-0}^{+0.057} (\mbox{syst.}) _{-0.032}^{+0} (\mbox{syst.}), $
where the first systematic error comes from the uncertainty of the continuum extrapolation while the second one comes from the uncertainty of the value of the coupling constant at the IRFP.}

\FullConference{31st International Symposium on Lattice Field Theory - LATTICE 2013\\
		July 29 - August 3, 2013\\
		Mainz, Germany}

\usepackage{amsmath}
\newcommand{\beq}{\begin{eqnarray}}
\newcommand{\eeq}{\end{eqnarray}}

\begin{document}

\section{Introduction}
We would like to give a novel renormalization scheme of the composite operators.
The basic idea of the renormalization condition of this scheme is that the renormalization factor is normalized by the tree level correlation function of the operators.
One of the practical advantage of this renormalization scheme is that any lattice boundary conditions can be applied. 

As an example, we consider the pseudo scalar operator for the SU($3$) gauge theory coupled to $N_f=12$ massless fermions.
It is known that the theory has the nontrivial IRFP~\cite{TPL}.
We obtain the mass anomalous dimension of the fermions from the renormalization factor of the pseudo scalar operator though
the partially conserved axial-currect (PCAC) relation.
The result of the mass anomalous dimension at the IRFP of the SU($3$) gauge theory coupled to $N_f=12$ massless fermion is
\beq
\gamma_m^*= 0.044 \hspace{5pt}_{-0.024}^{+0.025} (\mbox{stat.}) _{-0}^{+0.057} (\mbox{syst.}) _{-0.032}^{+0} (\mbox{syst.}), 
\eeq
where the first systematic error comes from the uncertainty of the continuum extrapolation while the second one comes from the uncertainty of the value of the coupling constant at the IRFP.
This talk is based on the paper~\cite{anomalous-dim}. 

\section{A novel renormalization scheme of the composite operators}
Let us consider some arbitrary composite operator ($H$).
If the theory is renormalizable, a nonperturbative renormalized coupling constants, {\it e.g.} the gauge coupling constant and the mass parameter of the fermions, can be defined by amplitudes of the observables.

The renormalization factor can be defined by the correlator of the bare operator ($H$),
\beq
C_{H}(t)&=& \sum_{\vec{x}} \langle H(t,\vec{x}) H(0,\vec{0}) \rangle.
\eeq
To obtain the finite renormalized value of the correlator, we introduce a nonperturbative renormalization factor ($Z_H$),
\beq
C_{H}^R (t) =Z_H^2 C_{H} (t).
\eeq
Here $C_{H}^R$ denotes a renormalized correlation function and it is finite.
On the other hand, the renormalization factor $Z_H$ and the nonperturbative bare correlation function diverge, and on the right hand side these divergences are canceled each other.

To define the renormalization scheme, we introduce the renormalization condition on the renormalized correlator, in which the renormalized correlator is equal to the tree level amplitude:
\beq
C_{H}^R(t)=C_{H}^{\mathrm{tree}} (t).\label{eq:rg-condition}
\eeq
The renormalization factor of the composite operator is thus defined by
\beq
Z_H \equiv \sqrt{ \frac{C_H^{\mathrm{tree}}(t)}{C_H (t)} },
\eeq
at some fixed propagation length ($t$).
Thus the factor $Z_H$ is normalized by the tree level value for each propagation length.
One of the practical advantages of our renormalization scheme is that a special lattice setup, such as a special boundary condition, is not needed.

\section{Step scaling function}
To obtain the scale dependence of the renormalization factor, we use the step scaling method~\cite{Capitani:1998mq}.
First, we introduce the discrete mass step scaling function from the factor $Z_P$.

In our renormalization scheme, the renormalization factor $Z_P$ has two independent scales, the propagation time ($t$) and lattice temporal size ($T$)
\footnote{We always fix the ratio of $T$ and $L$.}.
To see the scale dependence of the factor $Z_P$ we fix the ratio $r=t/T$.
Here $r$ takes a value $0< r \le1/2$ because of the periodic boundary condition on the lattice. 
We obtain the scale dependence of the factor $Z_P$ when we change both the physical propagation length and lattice size together.

\begin{figure}[h]
\begin{center}
   \includegraphics[width=4cm,clip]{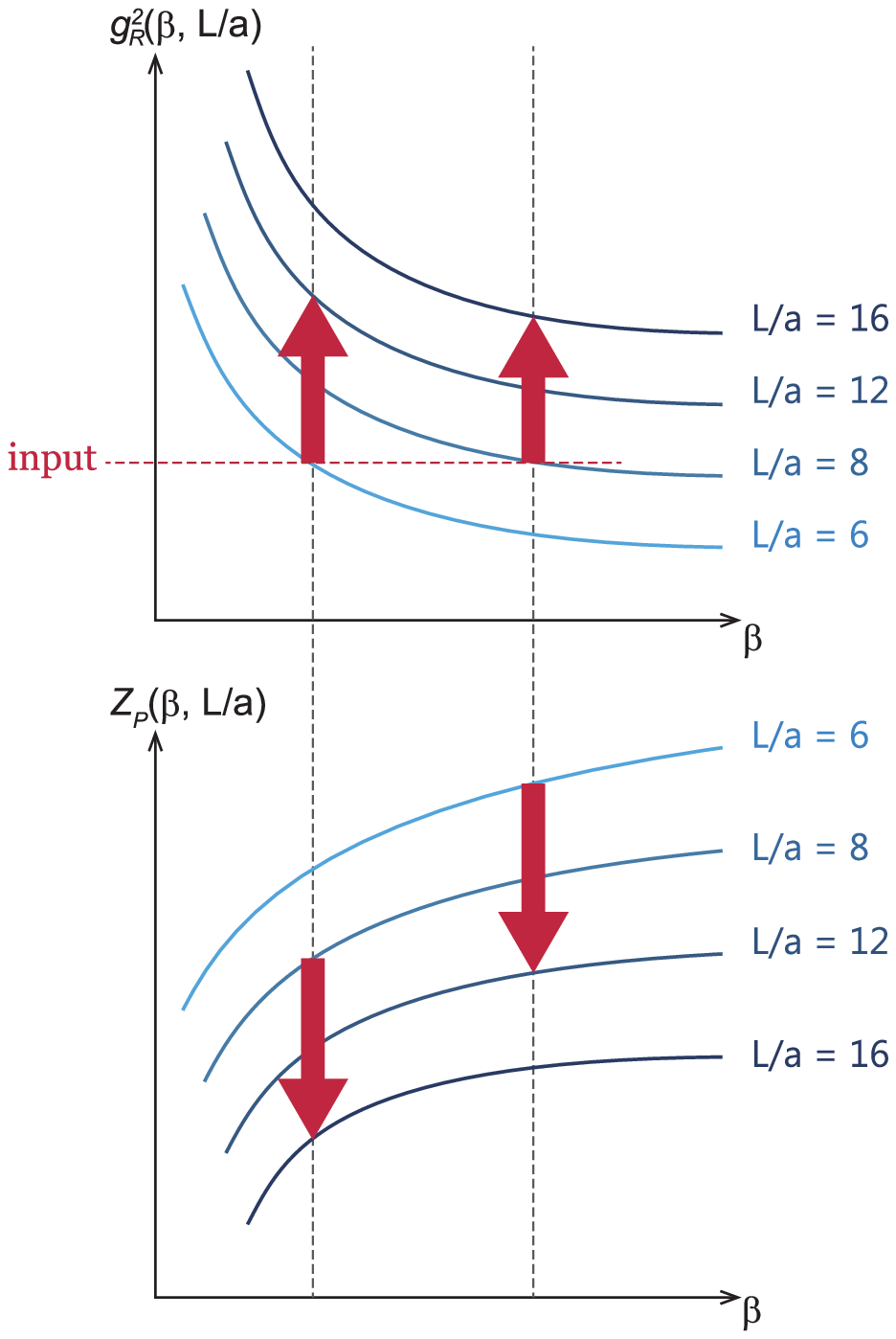}
  \caption{The strategies of the step scaling method for the coupling constant and the renormalization factor for the operator $P$. We measure the growth ratio of each quantity at fixed $\beta$.}
  \label{fig:strategy}
\end{center}
\end{figure}

A brief review of the strategy to obtain the mass anomalous dimension using the step scaling method is shown in Fig.~\ref{fig:strategy}.
The top panel shows the step scaling for the renormalized coupling.
To obtain the scale dependence of the renormalized coupling in a renormalization scheme, we measure the growth ratio of renormalized coupling when the lattice extent becomes $s$ times with fixed value of bare coupling constant.
And to obtain the scale dependence of the renormalization factor of a operator, we measure the growth ratio of the factor $Z_P$. 
The explicit definition of the mass step scaling function is given by
\beq
\Sigma_P (\beta , a/L; s)&=&  \frac{Z_P (\beta, a/sL)}{Z_P(\beta, a/L)}. \label{eq:disc-sigma}
\eeq
The mass step scaling function on the lattice includes the discretization error.
To remove it, we take the continuum limit ($a \rightarrow 0$) keeping the renormalized coupling ($u=g_R^2 (1/L)$) constant.
\beq
\sigma_P (u,s) &=& \left. \lim_{a \rightarrow 0} \Sigma_P (\beta, a/L;s) \right|_{u=const}.\label{eq:def-sigma}
\eeq

In the continuum limit, this mass step scaling function is related to the mass anomalous dimension. 
This relation becomes simple when the theory is conformal, and we can estimate the anomalous dimension at the fixed point with the following equation:
\beq
\gamma_m^*(u^*)=-\frac{\log |\sigma_P (u^*,s)|}{\log|s|}.
\eeq
Here $u^*$ denotes the fixed point coupling constant.

Note that in Eq.~(\ref{eq:def-sigma}), there is a freedom of the choice of the renormalization scheme concerning the input renormalized coupling constant. 
We use the set of the bare coupling constant and the lattice size to realize the input renormalized coupling constant ($u$) with a renormalization scheme for the gauge coupling constant.
The energy scale is defined by the input renormalized coupling, and the energy dependence of the mass step scaling function comes through the renormalized coupling constant.
We can use any combinations of the renormalization schemes for the renormalized coupling and the wave function renormalization.
Generally, the value of $\sigma_P (u)$ and the mass anomalous dimension depends on the choice of the renormalization schemes.
At the fixed point, although the value of $u^*$ depends on the renormalization scheme, the mass anomalous dimension is independent of the renormalization schemes of both the mass and the coupling constant.

\section{Simulation setup}
The gauge configurations are generated by the Hybrid Monte Carlo algorithm, and we use the Wilson gauge and the naive staggered fermion actions.
We introduce the twisted boundary conditions for $x, y$ directions and impose the usual periodic boundary condition for $z, t$ directions, which is the same setup with our previous work~\cite{TPL}.
Because of the twisted boundary conditions the fermion determinant is regularized even in the massless case, so that we carry out an exact massless simulation to generate these configurations.
The simulations are carried out with several lattice sizes ($L/a=6,8,10,12,16$ and $20$)
at the fixed point of the renormalized gauge coupling in the TPL scheme~\cite{TPL}.
In this simulation, we fix the ratio of temporal and spatial directions: $T/a=2L/a$

In the paper~\cite{TPL}, the IRFP is found at 
\beq
g_{\mathrm{TPL}}^{*2} = 2.69 \pm 0.14 (\mbox{stat.}) ^{+0}_{-0.16} (\mbox{sys.}),
\eeq
in the TPL scheme.
We use the tuned value of $\beta$ where the TPL coupling is the fixed point value for each $(L/a)^4$ lattices as shown in Table~\ref{table:beta-L}.
We generate the configurations using these parameters on $(L/a)^3 \times 2L/a$ lattices and neglect the possibility of induced scale violation coming from the change of lattice volume $(L/a)^4 \rightarrow 2(L/a)^4$ since we carefully take the continuum limit.
We measure the pseudo scalar correlator for $30,000$--$80,000$ trajectories for each $(\beta, L/a)$ combination.
We estimate the statistical error using bootstrap method.
\begin{table}[h]
\begin{center}
\begin{tabular}{|c|c|c|c|c|}
\hline
{} & {} &$g_{\mathrm{TPL}}^{2}=2.475$  & $g_{\mathrm{TPL}}^{2}=2.686$ & $g_{\mathrm{TPL}}^{2}=2.823$ \\     
\hline \hline
L/a & T/a & $\beta$ & $\beta$ & $\beta$  \\  
\hline
6   &  12 & 5.378          &  4.913           & 4.600         \\
8   &  16 & 5.796          &  5.414           & 5.181              \\
10   &  20 &  5.998          &  5.653           & 5.450            \\
12   &  24 &  6.121          &  5.786           & 5.588              \\
\hline
\end{tabular}
\caption{ The values of $\beta$ for each $L/a$ which give the TPL coupling constant at the IRFP.} \label{table:beta-L}
\end{center}
\end{table}

\section{Mass anomalous dimension at the IRFP}
Since we found that the data at $L/a=6$ has a large discretization error, we drop the data at $L/a=6$ from the continuum extrapolation of the step scaling function.
\begin{figure}[h]
\vspace{1cm}
\begin{center}
   \includegraphics[width=12cm]{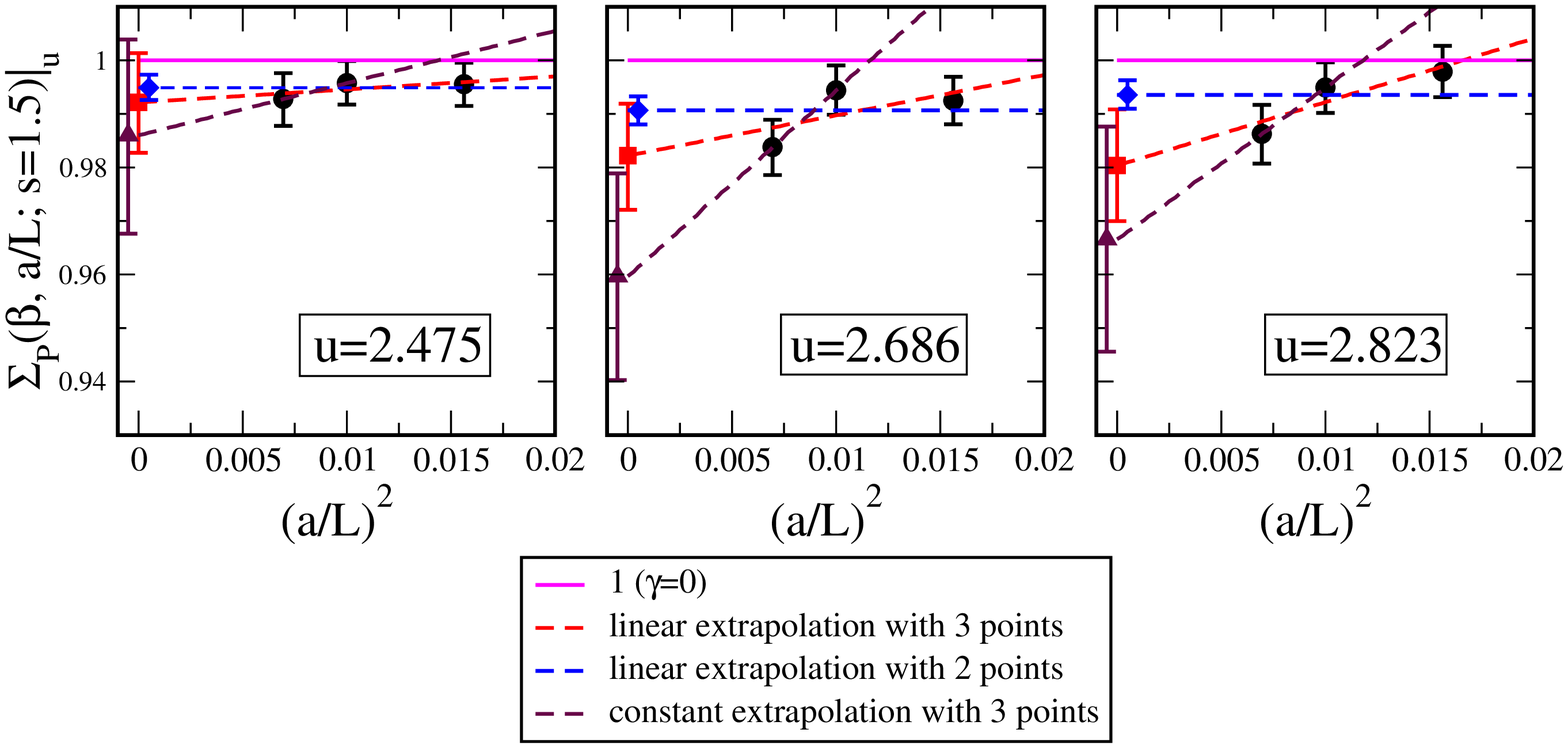}
 \caption{The continuum extrapolation of the mass step scaling function in $r=1/2$ scheme with $s=1.5$ without $L/a=6$ data. The solid line denotes the unity, where the anomalous dimension is zero. Each red and blue dashed lines denote the three-point linear and three-point constant extrapolations in $(a/L)^2$ respectively. The violet dashed line denotes the two-point linear function in $(a/L)^2$ using the finer two lattices.}
 \label{fig:cont-r-2-s-1.5-each-u}
\end{center}
\end{figure}
Figure~\ref{fig:cont-r-2-s-1.5-each-u} shows the finer three lattice data of the mass step scaling function ($\Sigma_P (\beta, a/L; s=1.5)$) and several continuum extrapolation functions.
As a central analysis, we take the three-point linear extrapolation in $(a/L)^2$, which is drawn in red dashed line for each panels.
The $u$ dependence is small and the result is consistent with each other.
We take the result for $u=2.686$ as a fixed point coupling as a central result.
The mass anomalous dimension of the central result is given 
$\gamma_m^*= 0.044 \hspace{3pt}_{-0.024}^{+0.025} (\mbox{stat.}) .$
The error denotes the statistical one, which is estimated by the bootstrap method.

We also carry out two different kinds of extrapolation.
One is the three-point constant extrapolation (the blue dashed line in Fig.~\ref{fig:cont-r-2-s-1.5-each-u}) and the other is the two-point linear extrapolation (the violet dashed line in Fig.~\ref{fig:cont-r-2-s-1.5-each-u}).
The smallest value of $\gamma_m^\ast$ is given by the three-point constant extrapolation for $u=2.475$, and the largest one is given by the two-point linear extrapolation for $u=2.686$.
Each value of $\gamma_m^\ast$ is $0.013$ and $0.102$ respectively.
We estimate the systematic uncertainties by taking the difference between the central value and smallest or largest value respectively.
Finally, we obtain the mass anomalous dimension 
\beq
\gamma_m^*&=& 0.044 \hspace{5pt}_{-0.024}^{+0.025} (\mbox{stat.})  \hspace{5pt}_{-0}^{+0.057} (\mbox{syst.})  \hspace{5pt}_{-0.032}^{+0} (\mbox{syst.}), \label{eq:final-gamma}
\eeq
where the first systematic error comes from the uncertainty of the continuum extrapolation while the second one comes from uncertainty of the fixed point coupling constant.

\section{Summary}
In this talk, we propose a novel renormalization scheme for the composite operators.
The renormalization factor is  normalized by the tree level correlation function for each composite operator.

We obtain the mass step scaling function at the IRFP of the SU($3$) $N_f=12$ massless fermion theory using this renormalization scheme for the pseudo scalar operator.
Our result of the mass anomalous dimension at the IRFP of this theory is given by
\beq
\gamma_m^*&=& 0.044 \hspace{5pt}_{-0.024}^{+0.025} (\mbox{stat.})  \hspace{5pt}_{-0}^{+0.057} (\mbox{syst.})  \hspace{5pt}_{-0.032}^{+0} (\mbox{syst.}).
\eeq

\begin{figure}[h]
\begin{center}
   \includegraphics[width=8cm]{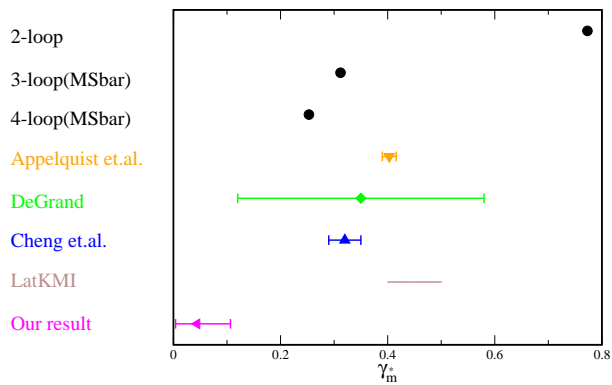}
 \caption{The comparison of the mass anomalous dimension at IRFP for several studies. From the top, the perturbative 2-loop result, 3-loop $\overline{{\mathrm{MS}}}$, 4-loop $\overline{{\mathrm{MS}}}$, the recent lattice results in the papers~\cite{Appelquist:2011dp}, \cite{DeGrand:2011cu},  \cite{Aoki:2012eq}, \cite{Cheng:2013eu} and our result. Note that in the papers~\cite{DeGrand:2011cu, Aoki:2012eq}  there is no `` $^*$ " on the gamma in their own papers. The value of the $\gamma_m^*$ of Ref.~\cite{Cheng:2013eu} is updated to $\gamma_m^*=0.25$ in Ref.~\cite{Hasenfratz:2013eka}.}
  \label{fig:comp-results}
\end{center}
\end{figure}

Finally, we show the values of the mass anomalous dimension in other literatures in Fig.~\ref{fig:comp-results}.
There are several discrepancies of the mass anomalous dimension although the mass anomalous dimension at the fixed point should be universal.
 We will give our updated result to obtain a conclusive value of the critical exponent at the IRFP in near future.

\section*{Acknowledgements}
The idea of this novel scheme was suggested to us by T.~Onogi.
Numerical simulation was carried out on
Hitachi SR16000 at YITP, Kyoto University,
NEC SX-8R at RCNP, Osaka University,
and Hitachi SR16000 and IBM System Blue Gene Solution at KEK 
under its Large-Scale Simulation Program
(No.~12/13-16), as well as on the GPU cluster at Osaka University.
We acknowledge Japan Lattice Data Grid for data
transfer and storage.
E.I. is supported in part by
Strategic Programs for Innovative Research (SPIRE) Field 5.


\begin{thebibliography}{99}
\bibitem{TPL} 
  E.~Itou,
  PTEP {\bf 2013}, no. 8, 083B01 (2013)
 
 
\bibitem{anomalous-dim} 
  E.~Itou,
  arXiv:1307.6645 [hep-lat].
  
\bibitem{Capitani:1998mq}
  S.~Capitani, M.~Luscher, R.~Sommer and H.~Wittig  [ALPHA Collaboration],
  Nucl.\ Phys.\  B {\bf 544} (1999) 669

  
    
\bibitem{Appelquist:2011dp}
  T.~Appelquist, G.~T.~Fleming, M.~F.~Lin, E.~T.~Neil and D.~A.~Schaich,
  Phys.\ Rev.\  D {\bf 84} (2011) 054501

\bibitem{DeGrand:2011cu} 
  T.~DeGrand,
  Phys.\ Rev.\ D {\bf 84}, 116901 (2011)
  
  


\bibitem{Aoki:2012eq} 
  Y.~Aoki, T.~Aoyama, M.~Kurachi, T.~Maskawa, K.~-i.~Nagai, H.~Ohki, A.~Shibata and K.~Yamawaki {\it et al.},
  Phys.\ Rev.\ D {\bf 86}, 054506 (2012)
  
  
\bibitem{Cheng:2013eu} 
  A.~Cheng, A.~Hasenfratz, G.~Petropoulos and D.~Schaich,
  JHEP {\bf 1307}, 061 (2013)


\bibitem{Hasenfratz:2013eka} 
  A.~Hasenfratz, A.~Cheng, G.~Petropoulos and D.~Schaich,
  arXiv:1310.1124 [hep-lat].




\end{thebibliography}
\end{document}